\documentclass[prd,twocolumn,nofootinbib,showpacs,amsmath,amssymb,floatfix]{revtex4}

\usepackage{graphicx}
\usepackage{dcolumn}
\usepackage{bm}
\usepackage{amsmath}
\usepackage{subfig}

\begin{document}

\title{Impact analysis of TOTEM data at the LHC: black disk limit exceeded.}

\author{A. Alkin}
\affiliation{%
Bogolyubov Institute
for Theoretical Physics, Metrologichna 14b, Kiev, UA-03680 , Ukraine.
}%
\author{O. Kovalenko }
\affiliation{%
Taras Shevchenko Kiev National University, Volodimirska 60, Kiev, UA-03101, Ukraine.
}%
\author{E. Martynov}
\affiliation{%
Bogolyubov Institute
for Theoretical Physics, Metrologichna 14b, Kiev, UA-03680 , Ukraine.
}%
\author{S.M. Troshin}
\affiliation{%
Institute for High Energy Physics, Protvino, Moscow Region, 142281, Russia.
}%

\date{\today}

\begin{abstract}
We discuss the profile of the impact--parameter dependent elastic scattering amplitude. Extraction of impact-parameter dependence from the dataset with inclusion of the experimental data on elastic scattering at the LHC energies helps to reveal the asymptotics of hadron interactions. Analysis of the data clearly indicates that the impact-parameter elastic scattering amplitude exceed the black disk limit at the LHC energy 7TeV and the inelastic overlap function reaches its maximum value at $b>0$.
\end{abstract}

\maketitle

\section{Introduction}
We show here that using the data on elastic differential cross-section can provide new information for the asymptotics of hadron scattering.
Of particular importance is the extraction of the impact--parameter dependent quantities from this experimental data including the recent measurement at LHC energies.

One of the attractive features of the impact parameter representation is a diagonalization of the unitarity equation for the elastic scattering amplitude $H(s,b)$, i.e.  at high energies
\begin{equation}\label{eq:unitarity}
 \mbox{Im} H(s,b)=|H(s,b)|^{2}+G_{inel}(s,b)
\end{equation}
with ${\cal O}(1/s)$ precision \cite{gold}. The term $|H(s,b)|^{2}$ is the elastic channel contribution, $G_{inel}(s,b)$ covers all the intermediate inelastic channels, and $b$ is an impact parameter of the colliding hadrons.

Information on $H(s,b)$, in particular, on $H(s,0)$, is necessary to select upper limit for this amplitude, namely, to know should this limit to be one half (it is the black disk limit) and correspond to the
maximum of the inelastic channel contribution to the elastic unitarity  with asymptotic ratio
\begin{equation}\label{bd}
  \sigma_{el}(s)/\sigma_{tot}(s)\to 1/2
\end{equation}
or it is equal  to unity and corresponds to a maximal value of the partial amplitudes allowed by unitarity resulting in the limit
\begin{equation}\label{eq:rd}
  \sigma_{el}(s)/\sigma_{tot}(s)\to 1
\end{equation}
at $s\to\infty$. Under assumption of  the limit 1/2 for the partial amplitude, the factor in the original Froissart-Martin
bound for the total cross-sections has been reduced by 2 \cite{mart}.
The bound reduced by factor of 4 for the total inelastic cross-section  has also been derived \cite{wu}.
Several asymptotic limits  have been treated in \cite{menon} in almost model-independent way,
but also for the forward scattering data only.

As well the Eq. (\ref{eq:unitarity})  is instrumental for the reconstruction of $G_{inel}(s,b)$\footnote{Though inelastic overlap function $G_{inel}(s,b)$ is not well suited for asymptotics studies} from the elastic scattering data\footnote{Cf. e.g. \cite{ama, fearn} for an earlier analysis of $G_{inel}(s,b)$ and \cite{dremin} for the most recent one.}.

The unitarity relation implies existence of the two scattering modes, designated as absorptive and reflective. Namely, the elastic scattering $\tilde S$-matrix element (related to the elastic scattering amplitude as $\tilde S(s,b)=1+2iH(s,b)$)
can be presented in the form
\[\tilde S(s,b)=\kappa(s,b)\exp[2i\delta(s,b)]\]
with the two real functions $\kappa(s,b)$ and $\delta(s,b)$. The function $\kappa$ ($0\leq \kappa \leq 1$) is an absorption factor\footnote{It has different meaning in the reflection region, as it will be discussed further.}, its value $\kappa=0$ corresponds to a complete absorption. At high enough energies the real part of the scattering amplitude can  be neglected,  allowing the substitution $H\to iH$. We consider this simplified case for the moment here. The choice of elastic scattering mode, namely, absorptive  or reflective, is governed by the phase $\delta(s,b)$. The common assumption is that  $\tilde S(s,b)\to 0$  at the fixed impact parameter $b$ and $s\to \infty$. It is called a black disk limit  and the elastic scattering   in this case is completely absorptive, i.e. it is just  a shadow of all the inelastic processes. This implies $\mbox{max\{Im}H(s,b)\}$=1/2.

There is  another possibility, namely, the function $\tilde S(s,b)\to -1$ when $b$ is fixed and $s\to \infty$, i.e.  $\kappa \to 1$ and $\delta \to \pi/2$. This case corresponds to a pure reflective scattering \cite{reflect}. The principal point is that the phase is non-zero, i.e. $\delta$ is equal to $\pi/2$ and $\mbox{max(Im}H(s,b))$=1.

We  discuss now the observable effects sensitive to the presence of the non--zero  phase. The most straightforward way is to extract impact-parameter dependent elastic scattering amplitude from the experimental data for the $pp$ and $\bar p p$ scattering.

\section{Impact analysis of the data}

Impact parameter analysis is performed following (with a minor modification) the method suggested by Amaldi and Schubert \cite{ama} for $pp$ scattering and applied by  Fearnley \cite{fearn} to $\bar pp$ scattering. Let us shortly describe how the amplitudes in impact parameter representation were extracted in \cite {ama, fearn} from the measured $d\sigma/dt$. We start with the relation between the impact $H(s,b)$ and standard $A(s,t)$ amplitudes ($b$ is given in {\it fm}).

\begin{equation}
\begin{array}{ll}
H(s,b)&=\frac{1}{8\pi s}\int\limits_{0}^{\infty}dq\, q J_{0}(qb/k_{1})A(s,t), \\
A(s,t)&=8\pi s\int\limits_{0}^{\infty}db\, b J_{0}(qb/k_{1})H(b,s), \quad t=-q^{2},
\end{array}
\end{equation}
$k_{1}=$ 0.1973269718 GeVfm. Normalization of $A(s,t)$ is the following (total cross section is measured in {\it mb})
\begin{equation}
\sigma_{t}=\frac{k_{2}}{s} Im A(s,0), \quad \frac{d\sigma}{dt}=\frac{k_{2}}{16\pi s^{2}}|A(s,t)|^{2}
\end{equation}
where $k_{2}$=0.389379338 mbGeV$^{-2}$.

To describe the data on $d\sigma/dt$ we used parameterizations of $A(t)\equiv A(s,t))$ (at fixed energy) modified from those in \cite {ama, fearn}.
\begin{equation}\label{eq:ampl}
\begin{array}{ll}
A(t)&=8\pi s\left \{ i\alpha (1-i\rho) \left (A_{1}e^{b_{1\alpha t/2}}+(1-A_{1})e^{b_{2\alpha t/2}}\right )\right .\\
&\left . -iA_{2}e^{b_{3}t/2}-A_{2}\rho (1-t/\tau)^{-4}\right \}
\end{array}
\end{equation}
where
\begin{equation}
\alpha=(1-i\rho)(\sigma_{t}/8\pi+A_{2}),
\end{equation}
$\rho$ and $\sigma_{t}$ are the real to imaginary part ratio of amplitude at $t=0$ and total cross section at the given energy. Parameters were fitted at each energy.

\begin{figure}[t]
    \centering
\includegraphics[scale=0.45]{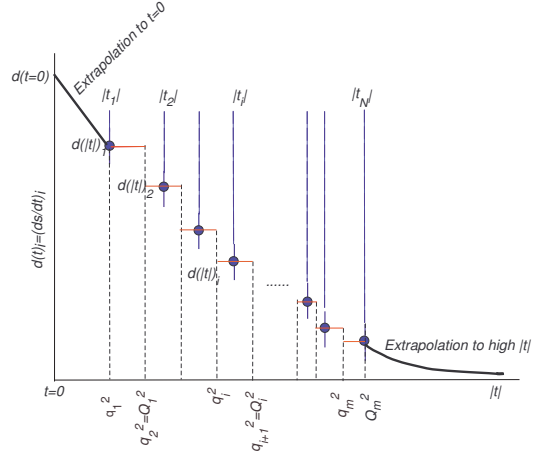}
\caption{Scheme of bins in the Eqs.~(\ref{eq:bins1}-\ref{eq:bins2})}
\label{fig:bins}
\end{figure}

{\bf Imaginary part of impact elastic scattering amplitude $H(s,b)$}\footnote{The profile function $\Gamma(s,b)=-2iH(s,b)$ has been extracted from the data in \cite{ama, fearn})} is calculated at each considered (and fixed) energy as
\begin{equation}\label{eq:bins1}
 \mbox{Im} H^{d}(b)=\frac{1}{8\pi s} \sum_{i = 1}^{N} \int_{q_{i}}^{Q_{i}} dq q
J_{0} \left( \frac {bq} {k_{1}} \right) I(q)_{i}
\end{equation}
where $N$ is number of points in the $d\sigma/dt$ data set at given energy,
\begin{equation}\label{eq:bins2}
I(q)_{i} = \sqrt{(16\pi s^{2}/k_{2})\left (d \sigma/d t \right )_{i} -(ReA)_{i}^{2} }
\end{equation}
and $(d\sigma/dt)_{i}$ is the experimental value of $d\sigma/dt$ measured at $t=t_{i}$ while $(\mbox{Re}A)_{i}$ is real part of the amplitude parameterized in the form (\ref{eq:ampl}) and calculated at $t=t_{i}$. The boundaries $q_{i}, Q_{i}$ of  $i$-th bin are defined as
\[ q_{i}^{2} = | t_{i} - t_{i-1}|/2,\]
\[ Q_{i}^{2} = | t_{i +1 } - t_{i}|/2.\]
Extrapolations to low and high $t$ were considered separately. Fig.~\ref{fig:bins} describes the entire scheme.

In the region $0\leq |t|\leq |t_{1}|$ the following extrapolation has been used
\begin{equation}
  A^{(low)}(t)=iA_{0}\exp(-B_{0}|t|)+ReA(t)/8\pi s
\end{equation}
where the real part of amplitude $ReA(t)$ is to be taken from the initial parametrization (\ref{eq:ampl}).
The constant $A_{0}$  can be found from the optical theorem
\begin{equation}
A_{0}=sk_{2}/\sigma_{t}.
\end{equation}
Here $\sigma_{t}$ is  the experimental value of  the total cross section at given energy. The slope $B_{0}$ is determined from the continuity condition at the first experimental point $t=t_{1}$
\begin{equation}
  \frac{d\sigma (t=t_{1})}{dt} =\frac{k_{2}}{16\pi s^{2}}|A^{(low)}(t_{1})|^{2},
\end{equation}
So, for lower $|t|$ values one can write ($t=-q^{2}$)
\begin{equation}
 \mbox{Im} H^{(low)}(b)= \frac{1}{ 16\pi s^{2} }
                       \int\limits_{0}^{|t_{1}|}dq\,
                       qJ_{0}\left( \frac {bq} {k_{1}} \right)ImA^{(low)}(t).
\end{equation}
Thus
\begin{equation}
 \mbox{Im} H(b)=  \mbox{Im} H^{(low)}(b)+  \mbox{Im} H^{(d)}(b)+  \mbox{Im} H^{(high)}(b)
\end{equation}
It can be shown that extrapolation to higher $|t|$, $ \mbox{Im} H^{(high)}(b)$ is negligible with any form of parameterizaion.

{\bf Uncertainty calculation.}   As the quantity under consideration depends on the data in a rather complicated way, uncertainties from the experimental points were propagated numerically by varying those within their respective limits (assuming the quoted uncertainty to be $\sigma$  interval) which produced a set of results for $\mbox{Im} H(b)$. The standard deviation of the resulting values of  $\mbox{Im}  H(b)$ at a given $b$ point was used as an uncertainty estimate.

{\bf Real part of $H(b)$ } is computed according to equation
 \begin{equation}
 \mbox{Re} H(b) = \frac{1}{8\pi s}
 \int \limits_{0}^{\infty}dq q J_{0} \left( \frac {bq} {k_{fm}} \right) \mbox{Re} A(q)
 \end{equation}
Standard error propagation formula can be used in this case. An error can be defined as
\begin{equation}
  \delta  \mbox{Re} H(b) = \sqrt{ \frac{ \partial\mbox{Re} H(b)}{ \partial p_{i}}
                                     \frac{\partial\mbox{Re} H(b)}{ \partial p_{j}}
                                      V_{ij} }
\end{equation}
Covariance matrix $V_{ij}$ for parameters  $p_{i}$ of the parametrization (\ref{eq:ampl}) was taken from minimization procedure (MINUIT).

\begin{figure}[t]
\begin{center}
\includegraphics[width=0.45\textwidth]{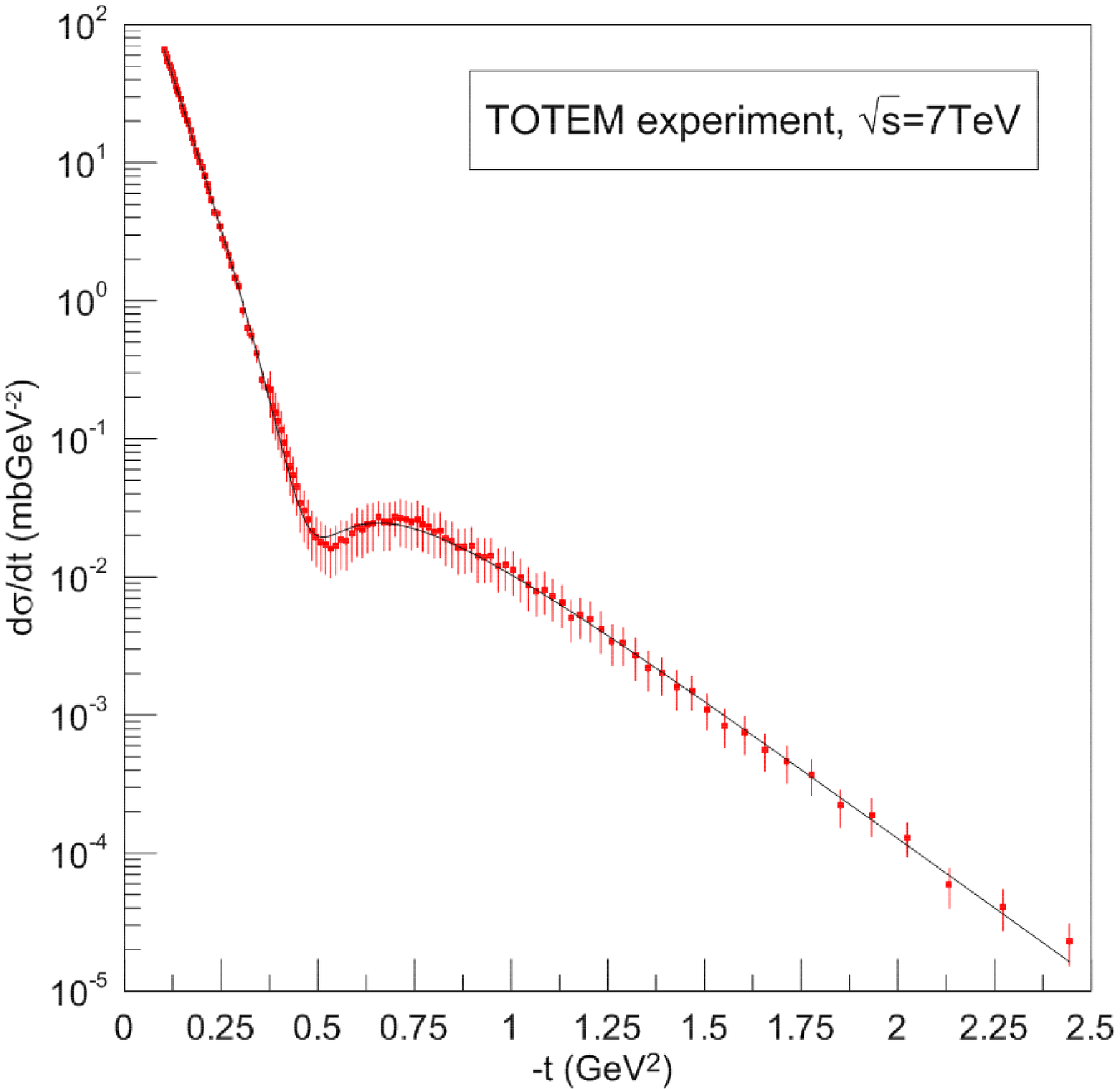}
\caption{Description of the TOTEM data at 7TeV with the amplitude parametrization (\ref{eq:ampl})}
\label{fig:dsdt}
\end{center}
\end{figure}

\begin{figure}[t]
\begin{center}
\includegraphics[width=0.45\textwidth]{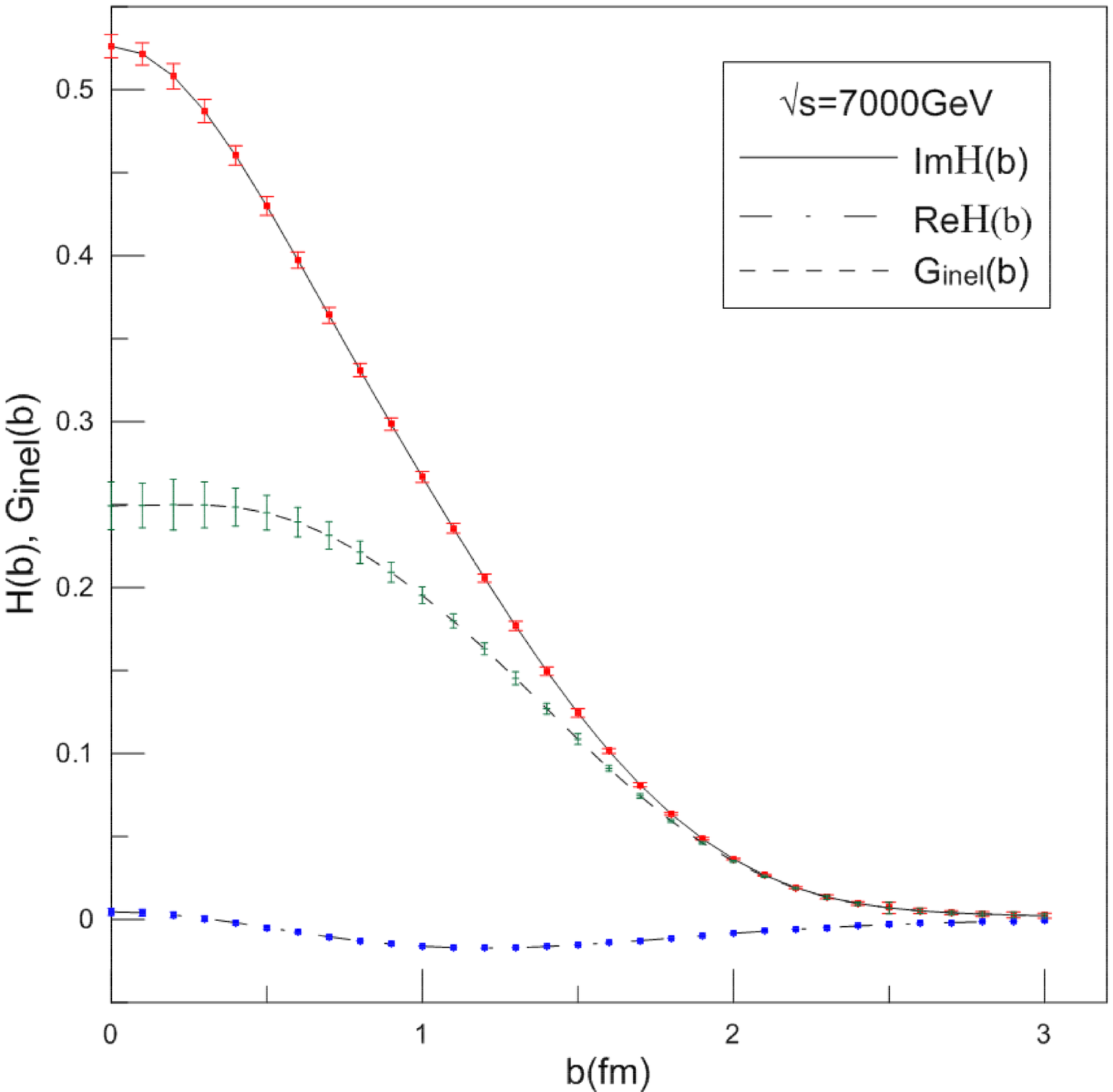}
\caption{Impact $pp$ amplitudes at 7TeV}
\label{fig:H,G-T}
\end{center}
\end{figure}

\begin{figure*}
\begin{center}
	\setcounter{subfigure}{0}
	\subfloat[a][$\mbox{Im} H(s,b)$]{
	\includegraphics[width=0.4\textwidth]{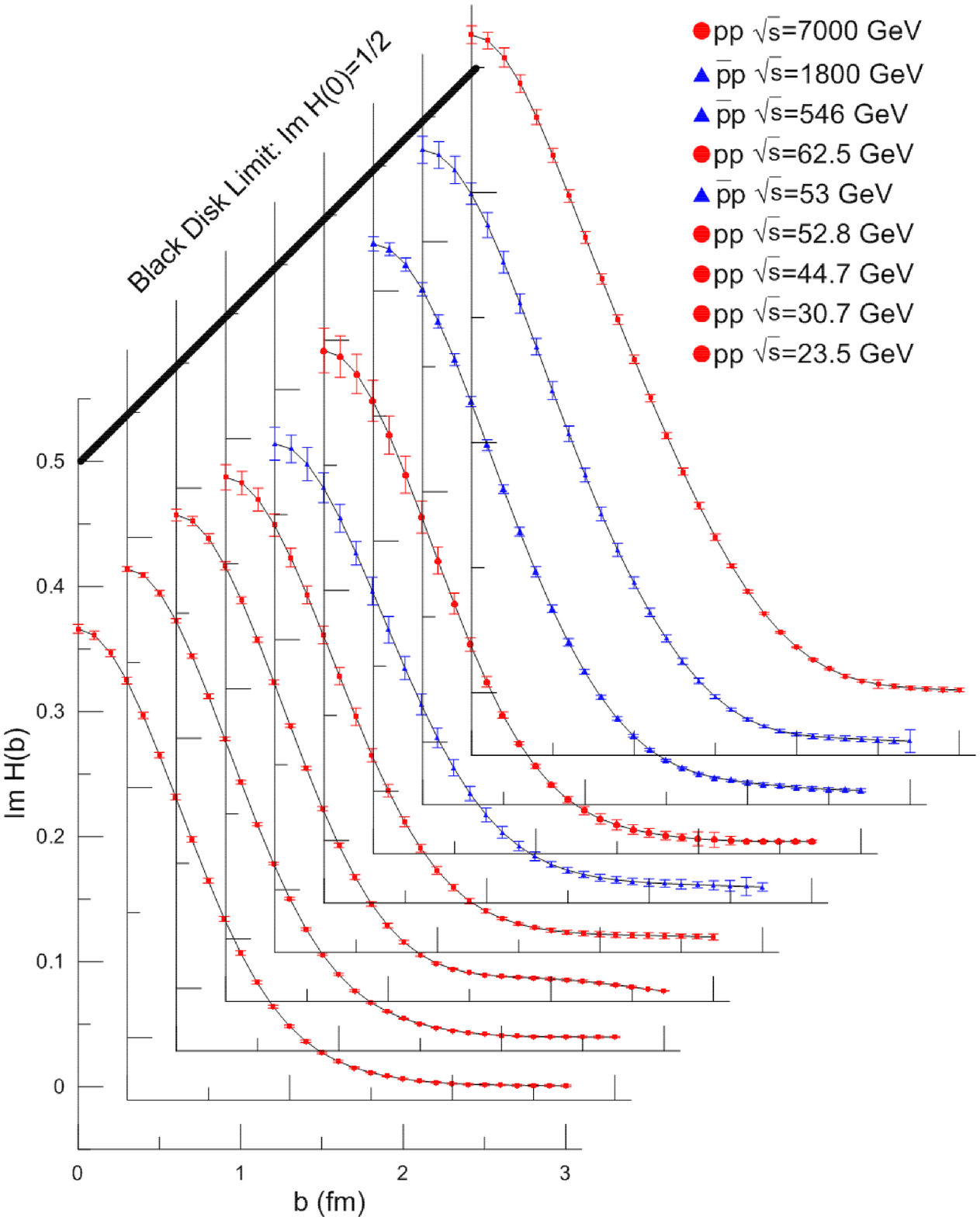}
	\label{fig:H-all}
	}
	\subfloat[a][G$_{inel}(s,b)$]{
	\includegraphics[width=0.4\textwidth]{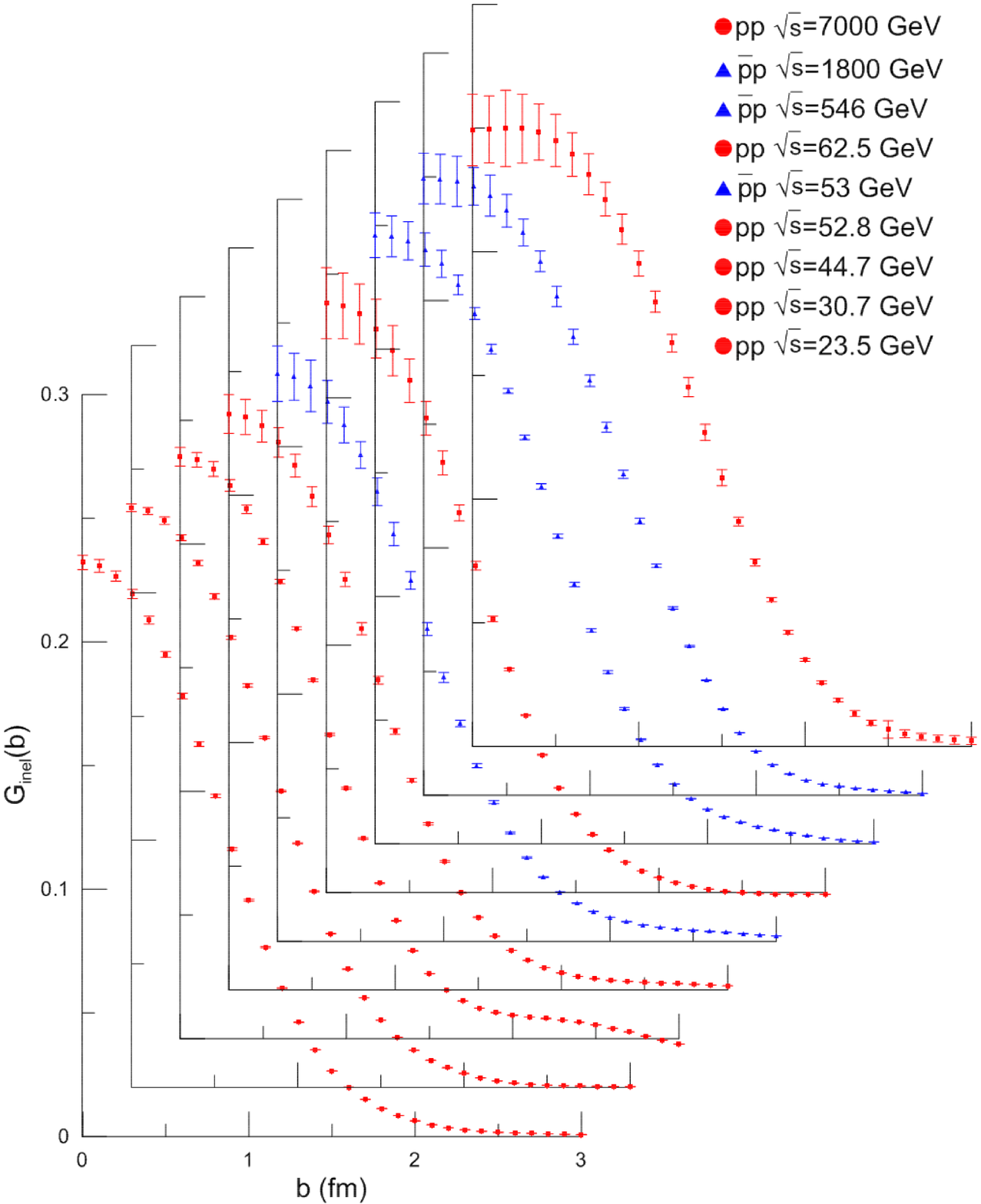}
	\label{fig:Gin-all}
	}
\end{center}
\caption{Imaginary parts of impact amplitudes and inelastic overlap functions of $pp$ and $\bar pp$ at various energies}
\end{figure*}

{\bf The results.} We have analyzed data on $pp$ elastic scattering at $\sqrt{s}=23.5, 30.7, 44.7, 52.8, 62.5, 7000 $ GeV \cite{ama, nagy, totem} and $\bar pp$ at $\sqrt{s}=53, 546, 1800 -1960$ GeV \cite{Breakst, Bozzo, Amos, Abe, D0}. The data at $|t|\geq 0.1$ GeV$^{2}$ were used for analysis. The main goal of our analysis is to extract $\mbox{Im} H(b) $ from the TOTEM data at $\sqrt{s}=7$TeV \cite{totem}. However in order to check the method we have applied it to older data to cross-check with \cite{ama, fearn}. We have found that our results for ISR, SPS and Tevatron  energies are compatible with those in \cite{ama, fearn}. Detailed explanation of our analysis will be presented in a separate paper. Here we demonstrate main results of our analysis, shown in the Figs.~\ref{fig:dsdt}-\ref{fig:Gin-all}. The Fig.~\ref{fig:dsdt} illustrates a quality ($\chi^{2}/df\approx 0.15$) of the TOTEM data description while the results of our impact analysis for $\mbox{Im}H(b), \mbox{Re}H(b)$ and $G_{inel}(b)$ at $\sqrt{s}=7TeV$ are presented in the Fig.~\ref{fig:H,G-T}. The most impressive fact is that $\mbox{Im} H(b)>1/2$ at small $b$. As was expected the $\mbox{Re} H(b)$ is quite small. In the Figs. \ref{fig:H-all}, \ref{fig:Gin-all} the evolution of $\mbox{Im}H(b)$ and $G_{inel}(b)$ is presented.

\begin{figure}[ht]
\begin{center}
\includegraphics[width=0.41\textwidth]{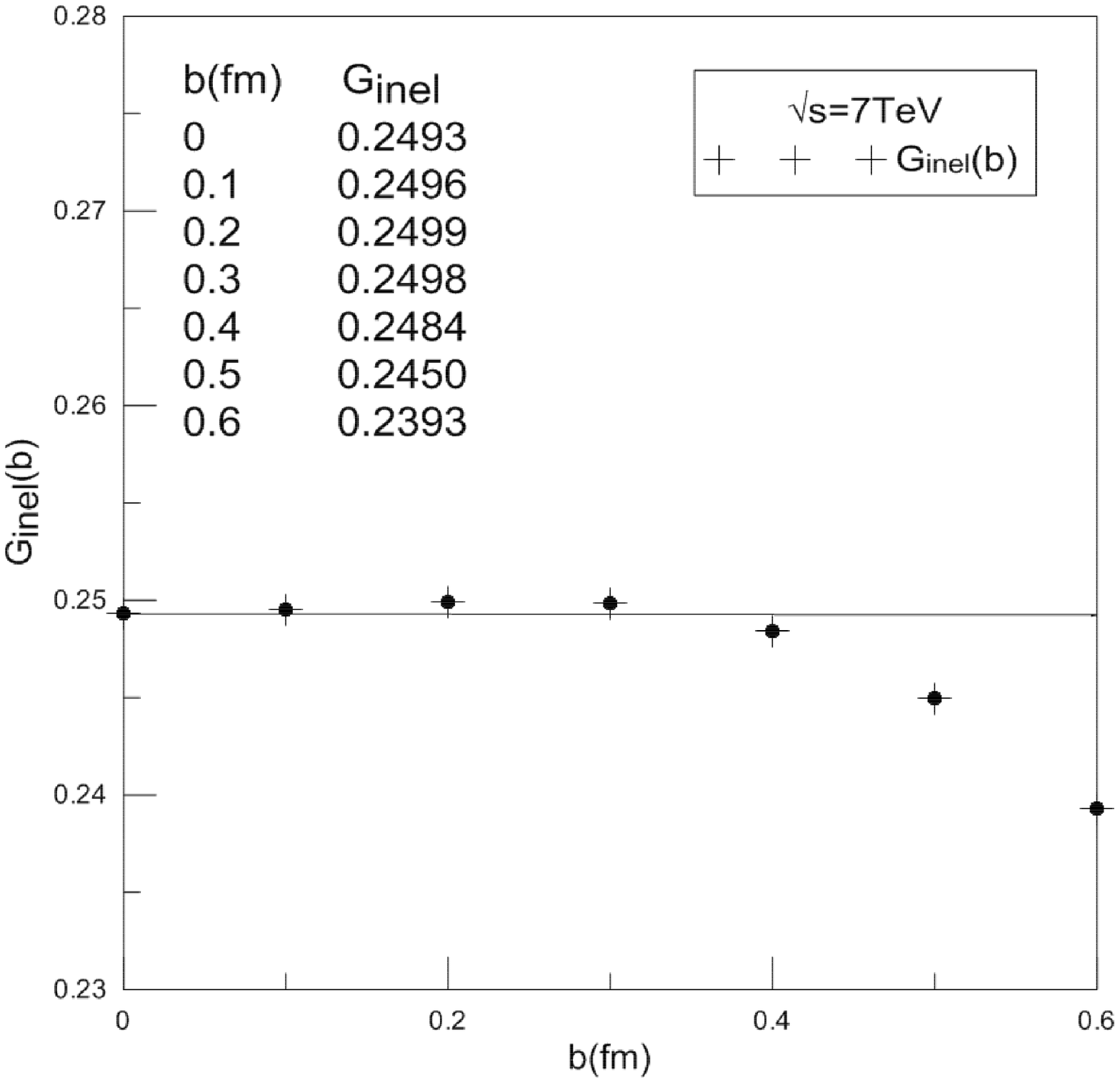}
\caption{The values of function $G_{inel}(b)$, extracted from the TOTEM data at small $b$.}
\label{fig:G-in-low-b}
\end{center}
\end{figure}

\section{Discussion of results}
Now we proceed to the qualitative implications of the above results. In the models where elastic scattering $\tilde S$-matrix element  $\tilde S(s,b)$  passes through zero to the negative values with increasing energy provide gradual
transition to the reflective scattering mode. This transition implies appearance of the phase $\delta =\pi/2$. The  solution of the equation $\tilde S(s,b)=0$ separates the regions of absorptive and reflective scattering and corresponds to the maximum value of $G_{inel}(s,b)=1/4$ since the derivative of $G_{inel}(s,b)$ has the form
\[
 \frac{\partial G_{inel}(s,b)}{\partial b}=\tilde S(s,b)\frac{\partial H(s,b)}{\partial b}
\]
and equals to zero at $\tilde S(s,b)=0$.  The derivative of the inelastic overlap function has the sign opposite to the sign of $\partial H(s,b)/\partial b$ in the region where $\tilde S(s,b)<0$ and the non-zero phase is,
therefore, responsible for the transformation of the central impact--parameter profile of the function $H(s,b)$ into a peripheral one of the inelastic overlap function $G_{inel}(s,b)$. It can also be easily seen by expressing the function $G_{inel}(s,b)$ as a product, i.e
\[
G_{inel}(s,b)=H(s,b)(1-H(s,b)).
\]
If $H(s,b)>1/2$ at high energy and small impact parameters, then the function $G_{inel}(s,b)$ will have a maximum value of 1/4 at the non-zero impact parameter value. We found some weak indication of such a maximum at $\sqrt{s}=7TeV$. In Fig.~\ref{fig:G-in-low-b} one can see that central values of the extracted $G_{inel}(s,b)$ data have a very shallow maximum at small $b$. The values of  $G_{inel}(s,b)$ at some $b$ point are also given in the figure. It would be interesting to see if this peak will be more pronounced at higher energies.

It should be noted that the derivative of the elastic overlap function has no sign-changing factor in front of $\partial H(s,b)/\partial b$, namely
\[
 \frac{\partial G_{el}(s,b)}{\partial b}=[1-\tilde S(s,b)]\frac{\partial H(s,b)}{\partial b}
\]
with  $1-\tilde S(s,b)$ being a non-negative at all values of $s$ and $b$.

The role of the non-zero phase in the high energy scattering is essential. In the presence of the non--zero phase at the LHC energies the reflective  scattering dominates at small impact parameters while inelastic processes are peripheral.  The albedo (coefficient of reflection) increases with energy at $s>s_0$  \cite{reflect}. The factor $\kappa(s,b)$ plays the role of albedo at $s>s_0$ and $b<R(s)$ and hence should be considered a reflective rather than absorption factor in this region.

Thus, the present analysis helps to understand which scattering mode is realized in asymptotics. Namely, assuming a monotonous energy dependence of the elastic scattering amplitude at the LHC energies and beyond one can conclude that reflective scattering mode is preferable on the base of this analysis which demonstrates that the elastic scattering amplitude exceeds the black disk limit at $\sqrt{s}=7$TeV. The near-future measurements of elastic scattering at the LHC energies $\sqrt{s}=10-13$TeV are very interesting and important for the confirmation or disproval of the above conclusion.

\section*{Acknowledgement}
We are grateful to N.E. Tyurin and G.M. Zinovjev for the many interesting discussions.

\end{document}